\documentclass[%
reprint,
 amsmath,amssymb,
showkeys,
]{revtex4-2}

\usepackage{graphicx}% Include figure files
\usepackage{dcolumn}% Align table columns on decimal point
\usepackage{bm}% bold math
\usepackage{enumitem} % Includes lists
\usepackage{amsmath}
\usepackage{color}
\usepackage{soul}

\def\e{\begin{equation}}
\def\f{\end{equation}}
\def\_#1{{\bf #1}}

\def\.{\cdot}

\def\@#1{_{\rm #1}}

\begin{document}

\preprint{APS/123-QED}

\title{Tunable Localization of Light Using Nested Invisible Metasurface Cavities}%
% \title{Nested Invisible Metasurface Cavities}% Force line breaks with \\
\thanks{Article published in Nanophotonics under DOI:\href{https://doi.org/10.1515/nanoph-2022-0549}{10.1515/nanoph-2022-0549}}%

\author{F.S. Cuesta}%
 \email{francisco.cuestasoto@aalto.fi}
\author{S. Kosulnikov}
\author{V.S. Asadchy}
%\author{M.S. Mirmoosa}
%\author{S.A. Tretyakov}%
\affiliation{%
 Department of Electronics and Nanoengineering, Aalto University,\\
 P.O. Box 15500, Aalto FI-00076, Finland    
}%

\date{\today}% It is always \today, today,

\begin{abstract}
An invisible cavity is an open resonant device that confines a localized field without producing any scattering outside of the device volume. By exploiting the scatter-less property of such device, it is possible to nest two invisible cavities, as the outer cavity would simply not notice the presence of the inner one, regardless of their relative position. As a result, the position of the inner cavity becomes a means to easily control the field localized inside the cavity and its quality factor. In this paper, we discuss the properties of nested invisible cavities as a simple method to achieve stronger localized fields and high tunable quality factor. Furthermore, we show that in optics, these cavities can be implemented using nanodisk-based dielectric metasurfaces that operate near their electric resonances.
 
\end{abstract}

\keywords{cavity, metasurface, tunable, field localization, quality factor}%Use showkeys class option if keyword
\maketitle

%\tableofcontents

\section{Introduction}
\begin{figure}[t]
    \centering
    \includegraphics[width=1\linewidth]{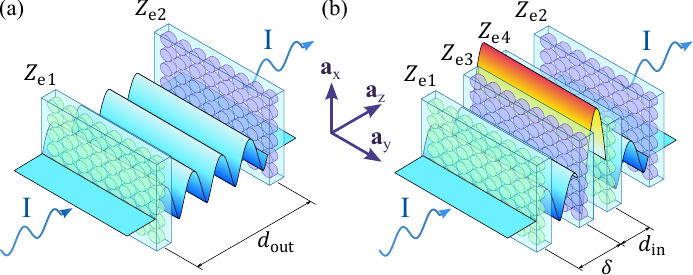}
    \caption{(a) Illustrations of a single invisible cavity formed by two parallel metasurfaces. Each metasurface consists of a uniform array of high-index disks.  Disks shown with light blue color possess electric resonance slightly above the resonant frequency of the cavity (capacitive metasurface), while dark blue disks have the same electric resonance below the resonant frequency (inductive metasurface). The complex magnitude of the electric field is shown across the cavity, being constant outside the cavity and standing-wave-like within it. 
    (b) Illustration of   nested invisible cavities formed by four metasurfaces.
    % is capable to support a standing wave inside, without producing any scattering outside. (b) Without the interference of such forward and backward scattering, it is possible to stack two or more of these cavities while remaining invisible. The inner cavity can move freely inside the outer cavity to control the properties of the localized field.
    }
    \label{fig:matryoshka_concept}
\end{figure}

Light localization plays an important role in modern physics and applications, particularly, in nanophotonics. It is the  basis for creating nonlinear devices, sensors, waveguides, resonators, photonic crystals, interferometers, optical fibers, lasers,  etc. With the advent of new nanofabrication technologies, optical systems enabling efficient and controllable light concentration at the nanoscale are on demand~\cite{John_1991_localization,Soref_2006_Future, Koenderink_2015_shrinking}. Foreseeable applications of such systems include   integrated circuitry for silicon photonics, optical computing, solar technologies, and bio-sensing.   

Recently, a possibility for light localization in a given volume with simultaneous partial or complete light scattering suppression from that volume
was discussed in multiple contexts, such as bound states in the continuum (BICs)~\cite{Marinica_2008_bound, Silveirinha_2014_trapping, Hsu_2016_bound, Azzam_2021_photonic, Plotnik_2011_bound, Zhen_2014_topological}, anapole modes~\cite{Miroshnichenko_2015_anapole, Wei_2016_excitation, Wu_2018_anapole, Yang_2019_anapole}, and virtual absorption~\cite{Baranov_2017_virtual,Longhi_2018_virtual,Marini_2020_open}. Such   nonradiating light concentration enables non-obstructive detection, enhancement and suppression of radiation, and energy storing. Probably, the most pronounced synergy between light localization and scattering suppression is achieved in the so-called invisible metasurface-based cavities~\cite{Cuesta_2020_non_scattering_cavities,Cuesta_2019_advanced}.  
These cavities allow to generate strong field maxima and deep minima inside them when illuminated by incident light, while they remain invisible to observers, both in the backward and forward directions. Such a regime is achieved due to the full suppression of light scattering from the cavity into outside space. It is in sharp contrast to  conventional Fabry-Perot resonators made of symmetric combinations of two partially transparent mirrors where
strong forward scattering is produced (resulting in a non-zero phase shift of the transmitted wave)~\cite{Sauleau_2005_fabry}. The invisible metasurface cavities have  an antisymmetric configuration of semi-transparent mirrors: one with weak  inductive and one with weak capacitive response. An illustration of such an invisible cavity is shown in Fig.~\ref{fig:matryoshka_concept}(a). Due to the antisymmetric (dual) and balanced electromagnetic response of the two sheets, the cavity remains invisible, being to some extent related to parity-time symmetric systems~\cite{Fleury_2014_negative, Radi_2016_parity, Monticone_2016_parity, Feng2017_nonhermitian, Zhao_2018_parity}. 
In this work, we consider invisible cavities formed by a pair of two electrically polarizable metasurfaces. It is worth mentioning that the present study can be also extended to invisible cavities formed by one electric and one magnetic, two magnetic~\cite{Cuesta_2020_non_scattering_cavities}, or two bianisotropic metasurfaces~\cite{Cuesta_2019_advanced}.

% The invisible cavity can be realized with different metasurface pairs: 
% from two sheets with electric response \cite{Cuesta_2020_non_scattering_cavities}, sheets with magnetic response, a combination of electric and magnetic response sheets, or from sheets with induced electric and magnetic currents \cite{Cuesta2018_invisible_cavities}. The concept of invisible cavities can be extended for bianisotropic sheets, where the additional degrees of freedom allows to retain invisibility in cavities with arbitrary length \cite{Cuesta_2019_advanced} and can be realized in optics using asymmetric unit-cells \cite{Alaee2015_biani_nanoparticle}. 

Due to this non-scattering nature, it was suggested that the invisible cavities can be ``nested''
%, one inside another, forming a ``matryoshka''-like configuration
[Fig.~\ref{fig:matryoshka_concept}(b)],  allowing further enhancing the field localization, still preserving the overall invisibility~\cite{Cuesta_2020_non_scattering_cavities,Cuesta_2019_advanced}.
However, implementation for such structures was conceptualized only for the microwave frequency range using metallic patterns. Simple scaling-down of this design to the optical frequency range is not possible since dissipation in metal appearing at high frequencies would strongly degrade the performance of the cavity (see Supplementary Section S1). 
% On the other hand, a nested cavity could be developed to operate in the optical range, with the use of low-loss dielectrics and metasurfaces with equivalent scattering properties.
%The proposed metasurface cavities were implemented at the microwave frequency range based on metallic patterns. However, this design cannot be simply scaled down to operate at the optical frequencies due to the fact that metals in that range become very lossy, which will inevitably compromise the quality factor of the cavity and its invisibility.

In this paper, we discuss the aspects of nesting invisible cavities and propose a low-loss optical implementation using metasurfaces with a high-permittivity nanodisks that operate near the electric resonance~\cite{Bohren_1983_absorption,Mishchenko_2002_scattering,Decker_2015_dielectric_Huygens, Kruk_2017_functional,Shankhwar_201718,Tzarouchis_2018_light,Bulgakov_2021}. 
%The resonant mode  is generated by the volumetric distribution of displacement currents~\cite{Bohren_1983_absorption,Mishchenko_2002_scattering,Decker_2015_highefficiency, Kruk_2017_functional,Tzarouchis_2018_light}. 
In order to provide required capacitive and inductive responses of the two metasurfaces, we slightly detune the dimensions of their unit-cell elements away from the   electric-dipole resonance condition. Furthermore, we design a nested configuration of two cavities that provides quadratic improvement of the quality factor as compared to that of a single cavity. Finally, we demonstrate that due to the invisibility property, the nested cavity enables a simple mechanism for tuning its quality factor (alternatively, the strength of field localization) in a   wide and continuous range of values.

\section{Nested Invisible Cavities}

\begin{figure}
    \centering
    \includegraphics[width=0.8\linewidth]{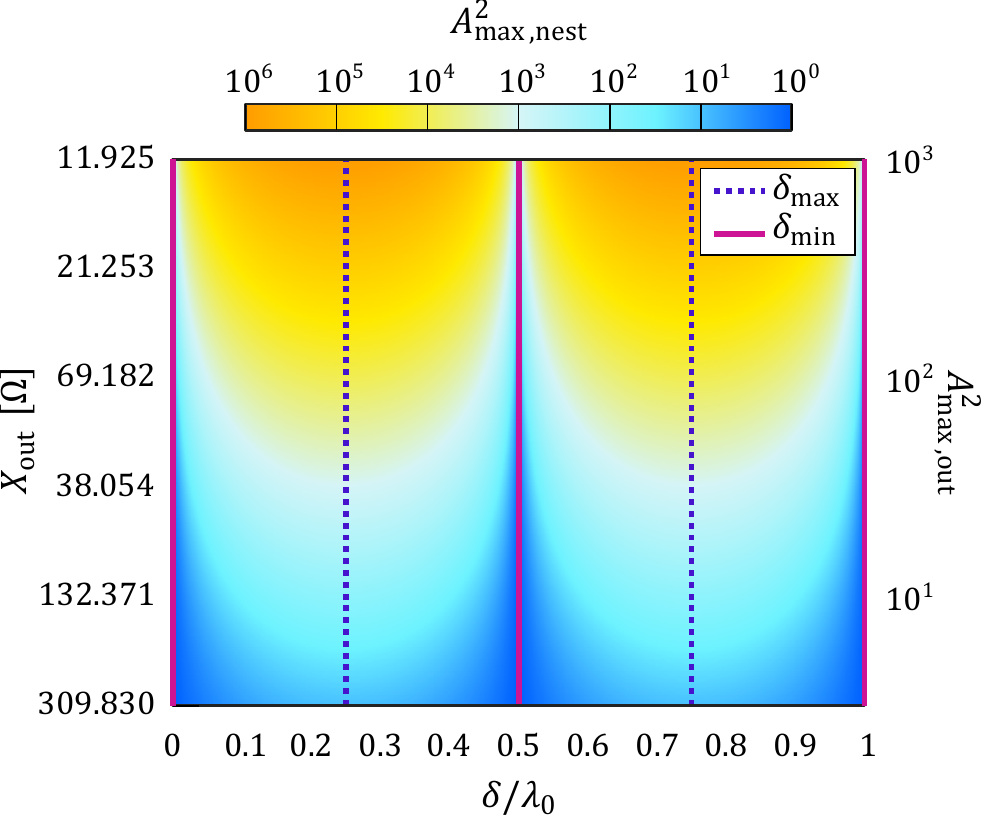}
    \caption{A nested invisible cavity can control the properties of standing waves by simply moving the inner resonator within the outer one. With the proper alignment, the maximum field localized in the inner cavity can be controlled in the range between $A\@{max,nest}^2=1$ (no standing wave) and a maximum of $A\@{max,nest}^2=A\@{max,out}^4$.  The reactances were chosen following an alternating pattern $X\@{in}=-X\@{out}$, with values that guarantee a field amplification of the outer cavity in the range between $A\@{max,out}^2\approx 3.16$ and $A\@{max,out}^2\approx 1000$.
    }
    \label{fig:swr_delta}
\end{figure}

The simplest nested invisible cavity consists of  two open cavity resonators, where one of the cavities is placed inside the other one, as portrayed in Fig.~\ref{fig:matryoshka_concept}(b). 
In this setup, since the inner cavity does not produce any scattering outside of it, its placement inside the outer cavity does not modify the  standing wave inside the inner cavity. 
The invisibility  of  the inner cavity implies the following~\cite{Cuesta_2020_non_scattering_cavities}: the complex grid impedances of the walls are complement in the form $Z\@{e3}=-Z\@{e4}=Z\@{in}$, and they are separated by a distance $d\@{in}=n\lambda_0/2$ [see Fig.~\ref{fig:matryoshka_concept}(b)], where $n$ is an integer and $\lambda_0$ is the operational free-space wavelength. The grid (or sheet) impedance  of a metasurface is the ratio between the averaged tangential  electric field  at the metasurface  and   the induced electric surface current density~\cite{Tretyakov_2003,Simovksi_2020}.
Likewise, invisibility   of the outer cavity occurs when
% The outer cavity can be either invisible, or it can have a symmetric Fabry-Perot design, as both resonators can provide the same standing wave \cite{Nahvi_2022_cavity}. Nevertheless, this work will consider an outer invisible cavity with 
$Z\@{e1}=-Z\@{e2}=Z\@{out}$ and $d\@{out}$ is proportional to half-wavelength at the operational frequency. By assuming that the metasurfaces have a uniaxial symmetry, with $z$ as the propagation direction of the incident wave, the standing wave of the outer cavity can be decomposed into its ``forward'' and ``backward'' components:
\begin{subequations}
\begin{equation}
    \_E\@{F,out}%=E\@{F,out}e^{-j k_0 z} \_x
    =E\@{I}\left(1-\dfrac{\eta_0}{2Z\@{out}}\right)e^{-j k_0 z}\_a\@{x},
\end{equation}
\begin{equation}
    \_E\@{B,out}%=E\@{B,out}e^{j k_0 z} \_x
    =E\@{I}\left(\dfrac{\eta_0}{2Z\@{out}}\right)e^{j k_0 z}\_a\@{x},
\end{equation}
\end{subequations}
where $k_0$ and $\eta_0$ are the free-space wavenumber and wave impedance, and
$E\@{I}$ is the amplitude of the incident electric field. Here, we assume the time harmonic convention in the form of $\exp (+j\omega t)$.
The field maximum of the standing wave $E\@{max, out}$ corresponds to the constructive combination of the forward and backward waves. The normalized field maximum  $A\@{max, out}$ is then given by
\begin{equation}
    A\@{max,out}=\dfrac{E\@{max,out}}{\vert E\@{I} \vert} =\dfrac{\eta_0+\left\vert\eta_0- 2Z\@{out}\right\vert}{\left\vert 2 Z\@{out}\right\vert}.
\end{equation}
The standing wave at the inner cavity is the result of the coherent illumination by the outer forward and backward waves (see Supplementary Sections S2 and S3). Therefore, the inner fields read
\begin{subequations}
\label{eq:inner_fields}
\begin{equation}
    \_E\@{F,nest}=\dfrac{E\@{I} \eta_0}{2Z\@{in}}\left[ 2\dfrac{Z\@{in}}{\eta_0}-\left(1+\dfrac{\eta_0 e^{2j k_0 \delta}}{2Z\@{out}-\eta_0} \right)\right]e^{-j k_0 (z-\delta)}\_a\@{x},
\end{equation}
\begin{equation}
    \_E\@{B,nest}=\dfrac{E\@{I} \eta_0 }{2Z\@{in}}\left[1 +\dfrac{\left(2Z\@{in}+\eta_0\right)e^{2j k_0 \delta}}{2Z\@{out}-\eta_0} \right]e^{j k_0 (z-\delta)}\_a\@{x}.
\end{equation}
\end{subequations}
It can be appreciated that both inner fields in Eqs.~\eqref{eq:inner_fields} are periodic functions that depend on the inner cavity position $\delta$. As a result, the amplitude of the inner standing wave can be tuned by   moving the inner cavity within the outer one, as presented in Fig.~\ref{fig:swr_delta}. 

The nested cavity concept offers a flexible design framework, depending on the combination of cavities with different $Z\@{in}$ and $Z\@{out}$, relative position of the inner cavity $\delta$, and the order of the sheets. This work will focus on a quite simple but attractive setup where both cavities have negligible losses  ($Z\@{in}=jX\@{in}$ and $Z\@{out}=jX\@{out}$), and their individual sheets have the same reactance values, but alternating between inductive and capacitive layers ($X\@{out}=-X\@{in}$). The first sheet of the system (the first one to be illuminated by the incident wave) was chosen arbitrarily to be inductive ($X\@{out}>0$), as it can be either inductive or capacitive without affecting the overall performance of the nested cavity. 
% This combination of resonant cavities does not only simplifies the fabrication of the nested device, as it only requires one kind of inductive and capacitive sheets; but also for their straight-forward performance in terms of field localization, as portrayed in Fig.~\ref{fig:swr_delta}. 
Thanks to the use of resonant cavities with the same field localization $A\@{max,out}=A\@{max,in}$, the combined field localization can reach a peak value of
\begin{equation}
    A\@{max,nest}= A\@{max,out}^2.\label{eq:sw_max_field}
\end{equation}
More interestingly, the alignment for such maximum field localization is independent of the used reactance $X\@{out}$, taking the form  $\delta\@{max}=\lambda_0/4 + n \lambda_0/2$. On the other hand, due to the nature of the standing wave, the field localized inside the inner cavity is minimized by displacing it towards $\delta\@{min}$, a quarter-wave away from $\delta\@{max}$, as shown in Fig.~\ref{fig:swr_delta}. In that case, there is no localized field, as the overall setup can be seen as two cascaded invisible cavities, and only a traveling wave can be seen throughout the inner region.

\section{Quality Factor}
In terms of frequency bandwidth, the combination of two nested invisible cavities can increase the quality factor of the combined structure, compared to the quality factor of each individual cavity. For the purpose of analyzing the quality factor, the transmitted wave amplitude was calculated analytically by satisfying  the boundary conditions for each sheet at an arbitrary frequency (see Supplementary Section S4).
%The quality factor, for this study, is defined as the ratio between the resonant frequency $f_0$ and the bandpass bandwidth ${\rm BW}$
%\begin{equation}
%    Q=\dfrac{f_0}{\rm BW}.
%\end{equation}
%The bandpass bandwidth is defined as the frequency range around $f_0$ where the power of the transmitted wave is at least half of the power coming from the incident wave. 
The quality factor, for this study, is defined as the ratio $Q=f_0/{\rm BW}$  between the resonant frequency $f_0$ and the bandpass bandwidth ${\rm BW}$, the region where the power of the transmitted wave is at least half of the power coming from the incident wave.
Due to the complexity of the analytical expression for the transmitted wave, the half-power passband was estimated numerically using an inner cavity with $d\@{in}=\lambda_0/2$, while the outer cavity measures  $d\@{out}=3\lambda_0/2$. The resonant frequency was arbitrarily defined as $f_0=1$~GHz. Both cavities use the same pair of grid impedances, with reactances fluctuating between $X\@{out}=11.925$~$\Omega$ ($A\@{max,out}^2\approx 1002$) and  $X\@{out}=291.617$~$\Omega$ ($ A\@{max,out}^2\approx 3.37$).
\begin{figure}
    \centering
    \includegraphics[width=0.8\linewidth]{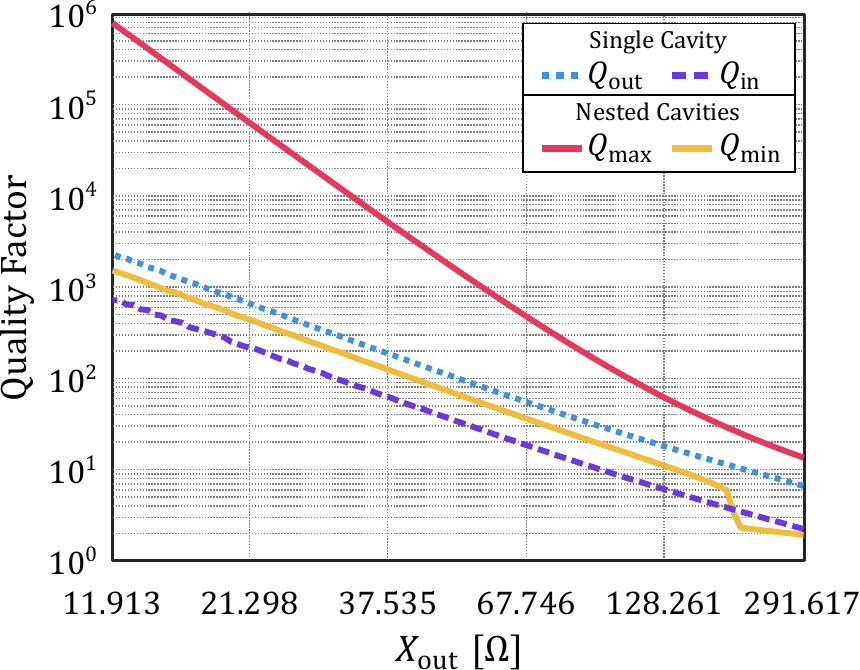}
    \caption{Quality factor of a nested cavity behaves similarly to its electric field maximum, being more resonant for lower values of the grid impedance, while the locations for maximum and minimum values are the same for the electric field maximum. Compared to the individual quality factor of the outer and the inner cavities, a nested cavity can offer a variable quality factor that can be similar to the individual cavities, or become more narrowband. The drop of $Q\@{min}$ for high $X\@{out}$ values is the result of neighbour four-layer resonant modes that extend the band-pass region. }
    \label{fig:q_factor_delta}
\end{figure}

%In comparison with the outer cavity, which has a higher quality factor due to its length, Fig.~\ref{fig:q_factor_delta} reveals that a nested invisible cavity can \red{modulate (what do you mean? you need to explain that by changind separation distance delta we can cover the space of quality factor values between the red and yellow curves. This is not obvious now from the text. )} its quality factor at will, becoming a strongly-resonant structure; or, if desired, it can obtain a broader bandwidth at the expenses of a weakened standing wave bounded in the inner resonator. 

Figure~\ref{fig:q_factor_delta} provides comparison of the quality factor of the nested cavity and the values achieved by the inner and outer cavities individually. For simplicity in numerical calculations, we consider non-dispersive sheets. Due to the resonant nature of the standing wave located at the inner cavity, it is expected that the alignment that produces strong field localization will also offer a high quality factor. Therefore, by placing the inner cavity at $\delta\@{max}$, the nested cavity becomes highly resonant, achieving its peak quality factor. Then, the quality factor can be controlled at will, decreasing down to its minimum value located at $\delta\@{min}$. 
In general terms, the lower boundary for the nested cavity quality factor is the one achieved for the inner cavity only. One exception is found for high reactance values of $X\@{out}$, where neighbour four-layer resonant modes combine with the main resonant mode, creating a broader band-pass region.

\section{Implementation with dielectric nanodisks}

\begin{figure*}
    \centering
    \includegraphics[width=0.9\linewidth]{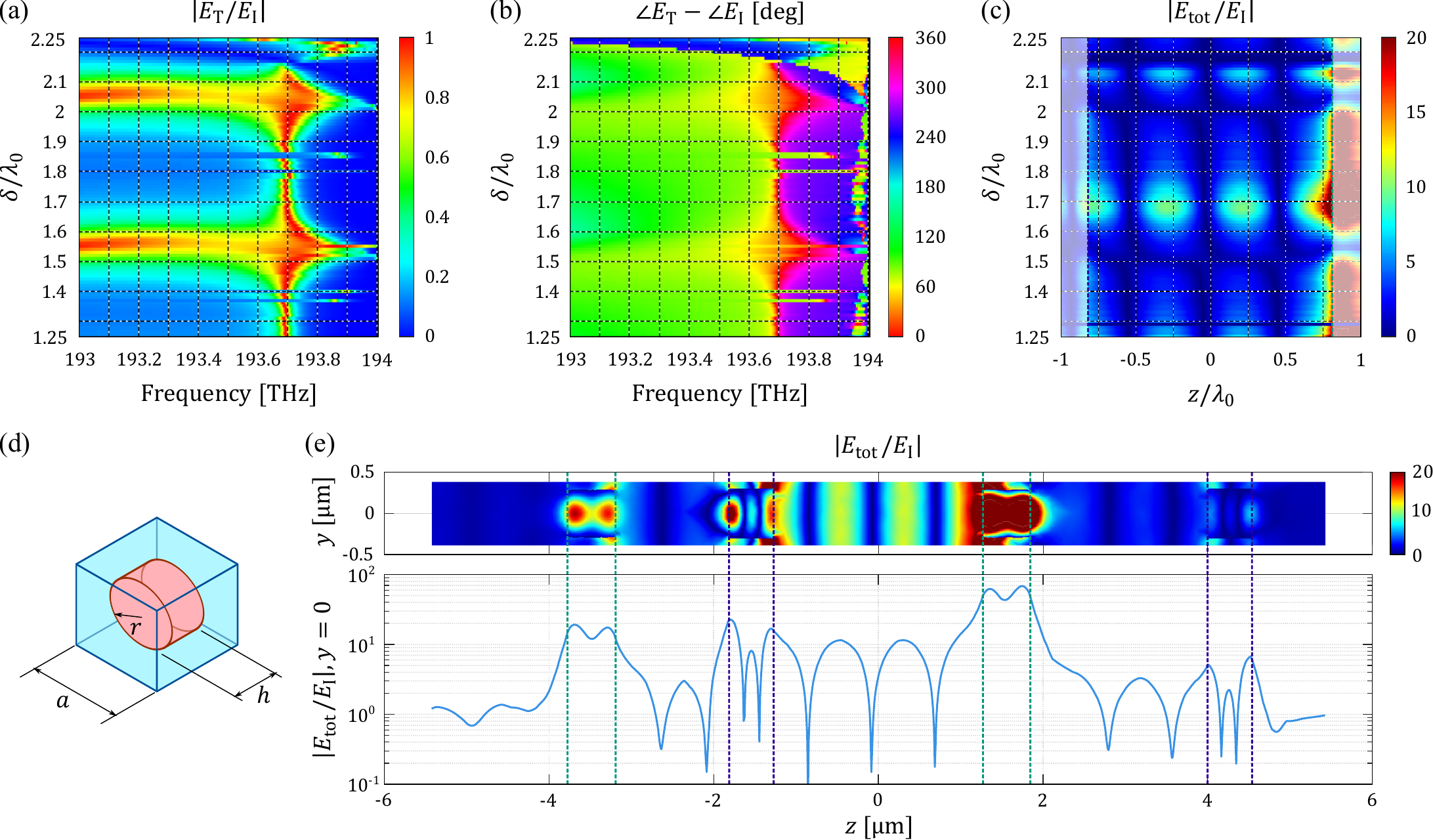}
    \caption{(a) After reducing the nanodisk dimensions to their two most significant digits, the nested cavity suffers a general drift of the operational frequency. (b) However, the nested cavity remains  invisible as the zero-phase difference between the incident and transmitted waves also shifts towards the operational frequency. (c) The nested cavity is able to control the field localized inside the inner cavity as a function of its relative position with the outer one $\delta$. Peak localization is achieved at  $\delta\@{max}=1.68\lambda_0$ with the peak electric field around ten times stronger than the incident one. The light-colored regions at the sides of the plot show the fields located inside the nanodisks. (d) The nested cavity was designed from dielectric nanodisk unit cells, with periodicity $a$, radius $r$, and height $h$. The optimized dimensions for the inductive and capacitive metasurfaces are presented in Table~\ref{tab:optic_metasurfaces_trimmed}. (e) A cross-section view of the total electric field distribution across the cavity ($\delta=1.68\lambda_0$ for this figure) reveals that the localized field in the inner cavity is not affected by the near fields of the  nanodisks   as long as sufficiently large distance between metasurfaces is chosen. The pairs of dashed lines show the extents and locations of metasurfaces.}
    \label{fig:optic_results}
\end{figure*}

As a proof of concept, a nested cavity is designed to operate in the near-IR range (with the design wavelength $\lambda_0=1550$~nm and frequency $f_0=193.41$~THz). The target field enhancement in the inner cavity, when it is aligned at $\delta\@{max}$, is chosen as nine times the incident field amplitude ($A\@{max,nest}=9$). This relatively small number was chosen deliberately to speed up  three-dimensional frequency-domain simulations (systems with high quality factors   generally require denser simulation meshes). It should be noted that, in principle, the field enhancement ratio is limited only by long simulation time or possible fabrication constraints.
According to Eq.~\eqref{eq:sw_max_field}, this enhancement can be achieved by nesting two individual cavities each of which provides maximum standing field amplitude of $A\@{max,out}=3 $. It can be found that these individual cavities can be formed    from inductive and capacitive metasurfaces with grid impedances $Z\@{e}=\pm j 0.375  \eta_0$. In optics, it is more convenient to describe metasurfaces  by their individual transmission ($\tilde{\tau}$) and reflection ($\tilde{\Gamma}$) coefficients, which are related to the grid impedance    $Z\@{e}$  as~\cite{Tretyakov_2003,Cuesta_2020_non_scattering_cavities}
\begin{equation}
    \begin{matrix}
       \tilde{\tau}=\dfrac{2 Z\@{e}}{2 Z\@{e}+\eta_0}, & \tilde{\Gamma}=-\dfrac{\eta_0}{2 Z\@{e}+\eta_0}.
    \end{matrix}
\end{equation}
For the inductive metasurfaces (\mbox{$Z\@{e}=j0.375 \eta_0$}), the scattering coefficients are written as \mbox{$\tilde{\tau}=0.6 \angle 53.13^{\circ}$}  and \mbox{$\tilde{\Gamma}=0.8 \angle 143.13^{\circ}$}. Likewise, the required capacitive metasurfaces (\mbox{$Z\@{e}=-j0.375  \eta_0$}) should have equivalent transmission \mbox{$\tilde{\tau}=0.6 \angle - 53.13^{\circ}$} and reflection \mbox{$\tilde{\Gamma}=0.8 \angle -143.13^{\circ}$} coefficients.
 
Implementation of such nested cavities in optics can be conveniently done using dielectric nanodisks  \cite{Decker_2015_dielectric_Huygens,Shankhwar_201718,Bulgakov_2021}, as depicted in  Fig.~\ref{fig:matryoshka_concept}. The meta-atom of Fig.~\ref{fig:optic_results}(d) represents a nanodisk made of silicon (refractive index $n=3.46$ near the design wavelength), surrounded by free space.  The nanodisk has radius $r$ (diameter $D=2r$) and height $h$. Its axis is oriented orthogonally to the metasurface plane.  Here, for clarity of the analysis, we omit   dielectric spacers between the individual metasurfaces that does not  qualitatively affects the presented results.
In order to reduce computational complexity during design, both metasurfaces have meta-atoms with the same square lattice with period $a=\lambda_0/2$. This particular value allows the excitation of electric dipole moments in the nanodisks, while preventing generation of diffraction orders for normally incident waves.  To design metasurfaces with small impedances $Z_{\rm e}= \pm j 0.375 \eta_0$, we first implement a resonant metasurface that emulates a perfect electric conductor (PEC) sheet with $Z_{\rm e}=0$ ($\tilde{\tau}=0$ and $\tilde{\Gamma}=-1$).  This can be done by choosing a combination of $r$ and $h$ that produces an electric dipole resonance at the design frequency~\cite{Tretyakov_2003}. It is worth mentioning that similarly one can realize invisible cavities formed by two magnetic~\cite{Cuesta2018_invisible_cavities} or two bianisotropic~\cite{Cuesta_2019_advanced} metasurfaces. While in the former case, one needs to exploit   the magnetic resonance of the nanodisks, the latter scenario can be reached with asymmetric nanodisks~\cite{Alaee2015_biani_nanoparticle}.
Next, we slightly scale up (scale down) the radius in order to depart from its dipole resonance and reach the desired response for the inductive (and capacitive) nanodisk layers (See Supplementary Material S5). 
% \red{In this case, the electric response is the result of exciting an anapole mode in both kind of nanodisk \cite{Liu_2020_multipole}.} 
% \blue{As in the limit case of invisible cavities, where it becomes a structure made of Perfect Electric Conductors (PEC) walls with a BIC inside it. Therefore the design started by choosing a combination of $r$ and $h$ that produces a dipole resonance at the design frequency, which behaves as PEC \cite{Tretyakov_2003}. }
Next, each kind of metasurface  was optimized separately using the built-in methods in CST (trust region algorithm for the inductive nanodisk and genetic algorithm for the capacitive one), with the required $\tilde{\tau}$ and $\tilde{\Gamma}$ as the optimization targets.
%Through optimization in CST, the best combination of parameters for each nanodisk were found to be for the inductive layers $r\@{ind}=0.19484 \lambda_0$ and $h\@{ind}=0.35102 \lambda_0$ [$\tilde{\tau}=0.3557 + j0.4791 (0.5967 \angle 53.4086^{\circ})$  and $\tilde{\Gamma}=-0.6403 + j0.4842 (0.8028 \angle 142.9031^{\circ})$]; while $r\@{cap}=0.1851 \lambda_0$ and $h\@{cap}=0.37205 \lambda_0$ [$\tilde{\tau}=0.3783 - j0.4202  (0.5654 \angle -48.0037^{\circ})$  and $\tilde{\Gamma}=-0.62 - j0.5443 (0.825 \angle 221.28^{\circ})$] for the capacitive ones.
In contrast to the inductive nanodisk, whose optimization was relatively easy (as its resonance frequency was far from the design frequency at $189.43$~THz), the capacitive nanodisk (with the resonance frequency of $193.65$~THz) required additional optimization steps, using CST built-in genetic algorithm. For instance, the capacitive nanodisk was optimized using first the outer cavity setup, with invisibility (unitary transmission with zero phase shift) at the design frequency as the optimization target. Then, the capacitive layers were tuned in the nested configuration by placing the inner cavity at $\delta\@{max}=1937.5$~nm (about 1.25 wavelengths), also with invisibility as the optimization target. This value of $\delta$ was purposely chosen to maximize the field localized inside the cavity and overall quality factor. Because of that, any small variations in the capacitive metasurfaces scattering would result in large variations in both  localized field and the overall transmission coefficient. During the single cavity and nested cavity optimizations, the inductive layers remained unchanged, as their individual transmission and reflection coefficients  were close to the optimization target from the beginning. Simultaneous optimization of both inductive and capacitive nanodisks was avoided to prevent other possible scenarios with different field localization strengths.
The separation  between the metasurfaces was chosen sufficiently large to prevent near-field coupling between contiguous layers. In that regards, the inner cavity was designed with $d\@{in}=2 \lambda_0$, while the outer cavity had $d\@{out}=5 \lambda_0$.

\begin{table}[]
    \centering
    \begin{tabular}{c|cc}
         & Inductive & Capacitive  \\
         \hline
        $D$ [nm] & 604.5 ($0.39 \lambda_0$) & 573.5 ($0.37 \lambda_0$)\\ 
        $h$ [nm] & 542.5 ($0.35 \lambda_0$) & 573.5 ($0.37 \lambda_0$)\\
        $a$ [nm] & 775 ($0.5 \lambda_0$) & 775 ($0.5 \lambda_0$)\\
        $\tilde{\Gamma}$ & $0.8015 \angle 142.82^{\circ}$ & $0.8586 \angle -141.79^{\circ}$ \\
        $\tilde{\tau}$ & $0.5980 \angle 53.82^{\circ}$ & $0.5127 \angle -52.34^{\circ}$       
%        $S_{11}$ & $-0.6713+j0.4718 (0.8205 \angle 144.9^{\circ})$ & $-0.1860-j0.4601 (0.4963 \angle 125.1^{\circ})$ \\
%        $S_{21}$ & $0.3557 + j0.4791 (0.5967 \angle 53.4086^{\circ})$ & $0.8049-j0.32489 (0.868 \angle -21.98^{\circ})$
    \end{tabular}
    \caption{Nanodisks dimensions (diameter $D=2r$, height $h$, and period $a$) of the inductive and capacitive metasurfaces with the most significant digit with respect to the design wavelength $\lambda_0=1550$~nm. It is also included the transmission $\tilde{\tau}$ and reflection $\tilde{\Gamma}$ coefficients produced by a single metasurface at the design frequency $f_0$.}
    \label{tab:optic_metasurfaces_trimmed}
\end{table}

After the optimization, the geometry of the nested cavity was slightly modified, considering fabrication constraints, as summarized in Table~\ref{tab:optic_metasurfaces_trimmed}. This modification is possible since the nested cavity is able to tolerate fabrication inaccuracies at the cost of shifting the operational frequency. In Supplementary Material Section~S6, we performed a tolerance study for the variation of nanodisk radii due to fabrication inaccuracies. For a typical 
variation of 4~nm~\cite{Levanon_2022},   a minor frequency shift in the operation of the cavity occurs at the level of 1\% of its operational frequency. Importantly, the transmission coefficient amplitude remains close to unity (with nearly zero phase) within the mentioned deviation range. The performance of the designed cavity is shown in Fig.~\ref{fig:optic_results} for different values of the separation distance $\delta$ between the inner and outer cavities. The most significant effect is a shift of the operational frequency away from the design frequency $f_0$ (where the nested cavity remains invisible) towards $f=193.695$~THz (a 0.2\% variation with respect to the design wavelength), as shown in Fig.~\ref{fig:optic_results}(a). Invisibility is verified by the results shown in Fig.~\ref{fig:optic_results}(b), where the phase difference between the incident and transmitted fields is close to zero around the operational frequency range. 
It can be noticed that the operational frequency remains almost stable regardless of the position of the inner cavity. However, for some discrete values of $\delta$, the resonant frequency of the nested cavity is moved. This effect takes place because the inner layers, which have thicknesses comparable to the wavelength and support Fabry-Perot-like modes \cite{Liu_2020_multipole}, suffer from destructive coherent illumination. By shifting the frequency, the single layers have a better suitable excitation for the standing wave inside the outer cavity.  In terms of field localization, Fig.~\ref{fig:optic_results}(c) reveals that this cavity produces field maxima at $\delta\@{max}=1.68\lambda_0$ close to $E\@{max}=9.9839 E\@{I}$, which is an increment of 10.93\% with respect to the target amplification. The strong field located in the right side of Fig.~\ref{fig:optic_results}(c) is the near field excited at one of the nanodisk interfaces. Field localization for alternative alignments at $\delta=1.25\lambda_0$ and $\delta=2.25\lambda_0$ do not reach the same levels found for $\delta\@{max}=1.68\lambda_0$ due to near-field interaction between consecutive layers, which feed one of the inner cavity walls with fields different from the expected standing wave. The frequency response in Figs.~\ref{fig:optic_results}(a)-(c) also demonstrates that the quality factor can be tuned by simply shifting the position of the inner cavity, as locating the inner cavity at the position of the maximum field localization also grants a high quality factor with a narrow pass-band region. A cross-section of the field distribution along the nested cavity, as portrayed in Fig.~\ref{fig:optic_results}(e), shows the presence of both standing waves (inner and outer). The additional space between metasurfaces has been proven effective, as the strong near fields around the nanodisks remains localized close to the metasurfaces.

\section{Conclusions}

In this work we presented and discussed the properties of a nested invisible cavities and their  implementations in the near infrared region. Each invisible cavity was formed by  inductive and capacitive metasurfaces. Both metasurfaces were designed from nanodisk meta-atoms with an electric-dipole resonance. The desired inductive and capacitive response was achieved by detuning the nanodisk dimensions to shift the dipole resonance. The resulting nested cavity offered a quadratic improvement of the quality factor achievable by the individual cavities. Similar property was observed in the field localization  inside the nested cavity: With the proper alignment, the inner field intensity increases by a power of two. Finally, we demonstrated that using an inner invisible cavity, which does not perturb the outer cavity, the resulting nested combination enables a simple means for tuning the quality factor and the field localized strength in a wide and continuous range of values. We note that it is possible to increase the number of nested cavities allowing further increase of localized field strength.

In practice, %nested tunability can be achieved depending of the cavity fabrication.
%Sergei: what is "nested tunability"?
if the cascaded metasurfaces are fabricated as four independent layers, the distance between the inner and outer cavities can be controlled using microelectromechanical systems (MEMS) nanopositioners \cite{Wang_2015_piezo,Maroufi_2021_nanopositioning}. Alternatively,   the optical distance $k_0\delta$ can be tuned through thermal modulation \cite{Weiss_2005_thermal_modulation}, or using laser pulses \cite{Guo_2004_fused,ORTYL_2005_diazo,Hofler_2006_oled,Lei_2016_phase}.
%  In this paper we presented the concept of nesting invisible cavities. An invisible cavity, as an open device, can be excited using coherent waves, with a standing wave controlled by such sources. In combination with its scatter-less property, an outer invisible cavity can be employed as the source of such coherent illumination, as the inner cavity would not interfere by creating additional scattering. As a result, the standing wave inside the inner cavity can be modulated using the relative position between the cavities, achieving strong field localization when the phase-difference between the contributors of the outer standing wave matches with the optimal phase illumination at the inner cavity.
The proposed  cavities can be used for 
cloaking sensors and obstacles, enhancement of
emission, tunable resonators for axion dark-matter haloscopes~\cite{mcallister_tunable_2018},
and creating exotic waveguides which are invisible in the direction orthogonal to their walls~\cite{Cuesta_2020_non_scattering_cavities}.

\noindent{{\bf Supplementary Material:} See Supplement Material at \href{https://doi.org/10.1515/nanoph-2022-0549}{https://doi.org/10.1515/nanoph-2022-0549} for supporting content.}

\noindent{{\bf Acknowledgements:} The authors would like to thank Prof. Sergei A. Tretyakov and Dr. Mohammad S. Mirmoosa for their useful comments and discussions related to this work.}

\noindent{{\bf Funding:} This work was supported in part by the Academy of Finland %(http://dx.doi.org/10.13039/501100002341)
under grant 330260 and by Nokia Foundation 
%(http://dx.doi.org/10.13039/501100004181)
under scholarship 20200224.}

\bibliography{References}

\end{document}